\documentclass[conference]{IEEEtran}
\IEEEoverridecommandlockouts
\usepackage{cite}
\usepackage{amsmath,amssymb,amsfonts}
\usepackage{algorithmic}

\usepackage{textcomp}
\usepackage{nameref}
\usepackage{graphicx}
\usepackage{xcolor}
\usepackage{tabularx}
\usepackage{url}
\usepackage{latexsym}
\usepackage{adjustbox}
\usepackage{longtable}
\usepackage{multirow}
\usepackage{threeparttable}
\usepackage{supertabular,booktabs}

\usepackage[utf8]{inputenc}
\usepackage{CJKutf8}
\usepackage{array}

\usepackage[colorlinks=true,citecolor=blue,linkcolor=blue,urlcolor=blue]{hyperref}

\def\BibTeX{{\rm B\kern-.05em{\sc i\kern-.025em b}\kern-.08em
    T\kern-.1667em\lower.7ex\hbox{E}\kern-.125emX}}

\begin{document}

\begin{CJK}{UTF8}{min}
\title{Facial Expression Recognition System Using DNN Accelerator with Multi-threading on FPGA
 \\
}

\author{\IEEEauthorblockN{Takuto Ando}
\IEEEauthorblockA{\textit{Electrical, Electronics Information Engineering Major} \\
\textit{National Institute of Technology, Oita College}\\
\textit{Advanced Course}\\
Oita, Japan \\
aes2301@oita.kosen-ac.jp}

\and

\IEEEauthorblockN{Yusuke Inoue}
\IEEEauthorblockA{\textit{Department of Information Engineering} \\
\textit{National Institute of Technology, Oita College}\\
Oita, Japan \\
y-inoue@oita-ct.ac.jp}
}

\maketitle

\begin{abstract}
    \label{abstract}
    In this paper, we implement a stand-alone facial expression recognition system on an SoC FPGA with multi-threading using a Deep learning Processor Unit (DPU).
    The system consists of two steps: one for face detection step and one for facial expression recognition.
    In the previous work, the Haar Cascade detector was run on a CPU in the face detection step due to FPGA resource limitations,
    but this detector is less accurate for profile and variable illumination condition images.
    Moreover, the previous work used a dedicated circuit accelerator, so running a second DNN inference for face detection on the FPGA would require the addition of a new accelerator.
    As an alternative to this approach, we run the two inferences by DNN on a DPU, which is a general-purpose CNN accelerator of the systolic array type.
    Our method for face detection using DenseBox and facial expression recognition using CNN on the same DPU enables the efficient use of FPGA resources while maintaining a small circuit size. 
    We also developed a multi-threading technique that improves the overall throughput while increasing the DPU utilization efficiency. 
    With this approach, we achieved an overall system throughput of 25 FPS and a throughput per power consumption of 2.4 times.

    \end{abstract}


\begin{figure*}[t]
  \centerline{\includegraphics[width=.95\textwidth]{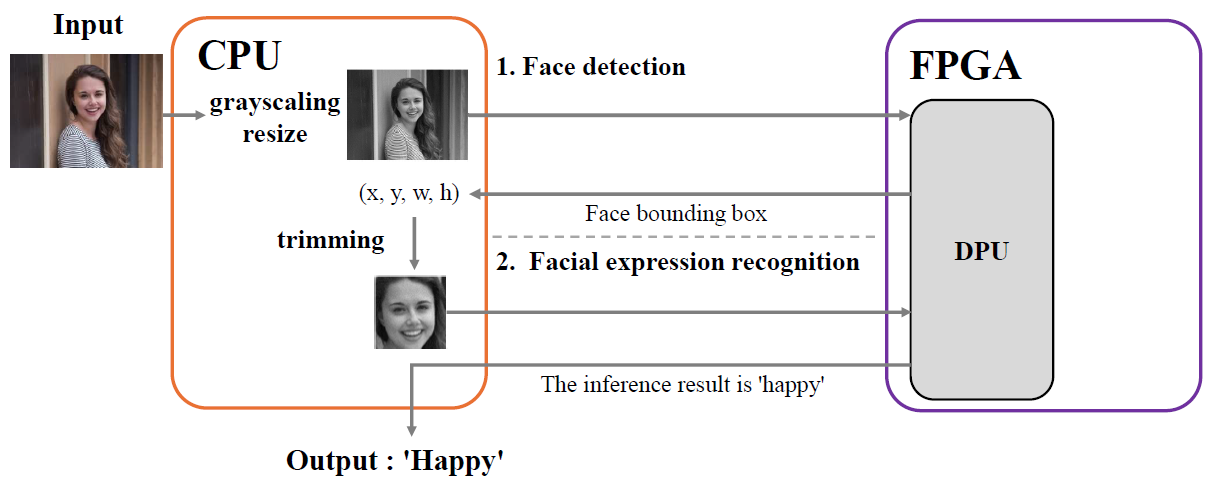}}
  \caption{System configuration of proposed method}
  \label{fig:arch}
\end{figure*}
\section{Introduction}
\label{introduction}
As the development of the robotics field continues to expand, robots are being utilized in an increasing variety of situations. 
For example, robots are now being implemented not only in industrial applications but also in service fields such as care, education, and entertainment, where the ultimate aim is to have them coexist with humans. 
Facial expressions play a key role in human-robot interactions\nobreak\cite{CHEN201849} as they are an effective non-verbal means of recognition\nobreak\cite{tian2001recognizing}, 
so the development of facial expression recognition technology is important for robots' understanding of emotions and communication.
Facial expression recognition technology has been applied in pet-type robots and medical robots,\nobreak\cite{Li2019TomPR,arriaga2017real},
and related methods utilizing  machine learning (such as support vector machines) have been proposed\nobreak\cite{ghimire2017facial}. 
Deep neural network~(DNN)-based methods\nobreak\cite{efficientCNN2021,shao2019three} have become mainstream and have achieved significant improvements in recognition accuracy over conventional methods.
Facial expression recognition by image processing extracts the region of interest of a face as a pre-processing step using a face detector such as the Cascade detector and then performs facial expression recognition within the area of interest.
Previous face detection methods have used the Viola-Jones Haar Cascade classification algorithm\nobreak\cite{inproceedings}.
DNN-based methods are also becoming more prevalent for face detection, and highly accurate detection methods using R-CNN\nobreak\cite{sun2018face} and YOLO\nobreak\cite{10.1007/s00371-020-01831-7} have been proposed.
Thus, we can expect face detection and facial expression recognition system components to achieve accuracy levels suitable for real-world application by implementing them in DNN-based methods.

Computation on a general-purpose graphics processing unit~(GPGPU) is typically used for DNN inference, as it offers a high computing performance.
However, it is not suitable for battery-powered robots due to its significant power consumption during inference.
For this reason, field programmable gate array (FPGA) execution is attracting attention as an alternative to GPGPU in embedded systems such as robots, 
as it can deliver the necessary performance to accelerate DNN inference processing while consuming less power.



When performing DNN inference on an FPGA, a CNN accelerator can be implemented in the circuit to enable fast inference processing.
When a facial expression recognition system is implemented on an FPGA, two inference processes using DNNs, namely, face detection and facial expression recognition, are accelerated. 
To perform different inferences on the same FPGA in this case, a dedicated accelerator on the FPGA is required for each inference. 
However, increasing the number of accelerators leads to higher consumption of FPGA resources and an increase in circuit size.
Therefore, by using the same accelerator to perform multiple inference processes, it could be possible to achieve a high processing performance while maintaining a small circuit size. 
In response to these challenges, we implement a facial expression recognition system using a DPU and propose a system that incorporates multi-threading to increase the overall throughput while increasing the efficiency of DPU utilization. 
Our main contributions are as follows.

\begin{itemize}
  \item We show that, in the pre-processing step for facial expression recognition, our DNN inference with DPU achieves a higher accuracy and shorter latency than the Haar Cascade detector.
  \item We demonstrate that utilizing multi-threading on the DPU improves the utilization rate of the DPU, thereby increasing the overall throughput.
  \item We clarify the optimal size and operating frequency of the DPU for the facial expression recognition system and identify the optimal DPU settings, taking into consideration the fact that CPU processing can create a bottleneck.
\end{itemize}


We organize the remainder of the paper as follows. 
Section~\ref{related_work} introduces two relevant previous works  (face detection using a Haar Cascade detector in a  SoC FPGA, and a hardware acceleration using DPU).
In Section~\ref{fer} we propose our facial expression recognition system using DPU.
Section~\ref{ex_re} presents the experimental evaluation we conducted, and Section~\ref{discussion} provides a discussion on the results. 
We conclude in Section~\ref{conclusion} with a brief summary and mention of future work.


\section{related work}
\label{related_work}
\subsection{Facial expression recognition system on FPGA}
Several works have investigated the implementation of facial expression recognition on FPGAs.
Most of these approaches are based on the assumption that face images can be accurately acquired through face detection. Consequently, only facial expression recognition methods have been proposed\nobreak\cite{KIM2021resource,phan2019fpga}.
In contrast, Vinh et al. developed a standalone system that performs facial expression recognition on the detected faces after conducting face detection on a SoC FPGA\nobreak\cite{vinh2019facial}.
In their facial expression recognition system, face detection is run on an ARM CPU, while facial expression recognition is run on an FPGA.
The facial expression recognition achieved high accuracy because the inference by DNN is run on the FPGA.
However, due to FPGA resource limitations, the OpenCV Haar Cascade detector is utilized for face detection.
Although the Haar Cascade detector can be run on an embedded CPU with a low computing performance, 
it can only detect faces directly facing the camera and has lower detection accuracy when the face is at an angle or sideways. 
Moreover, its performance is affected when the lighting conditions are not consistent. 
Consequently, face detection using non-DNN cannot accurately detect faces in real-world scenarios, thus reducing the accuracy of facial expression recognition.

When the hardware configuration above is utilized, it is necessary to add  a dedicated face detection accelerator to carry out face detection through the DNN inference. 
This expansion results in a substantial increase in circuit size, which poses a challenge due to the limited resources of the FPGA.
In our work, we propose a system that is capable of running on the DPU, a general-purpose CNN accelerator, to conduct face detection using DNN inference.
The DPU, provided by Xilinx, is an accelerator that can support various CNN inference processes within the same unit.
Thanks to utilizing this DPU, we can successfully implement our facial expression recognition system while keeping the increase in circuit size to a minimum.



\subsection{Hardware acceleration with DPU on FPGA}
Several works have investigated using the DPU to accelerate DNN on FPGAs. 
Li et al. proposed an improved YOLO model for optical remote sensing images called receptive field attention YOLO (RFA-YOLO) \nobreak\cite{LI2023edge}
and ran it on the Xilinx ZCU104 board using DPU to achieve high detection accuracy and reduce processing time. 
They conducted comparative experiments on three hardware platforms (CPU (Intel i7670), GPU (NVIDIA Tesla V100), and FPGA) and reported that the inference throughput and average power consumption of the inference run on the FPGA were 27.97$\,\mathrm{FPS}$ and 15.82$\,\mathrm{W}$, respectively. 
The power consumption was reduced by 62.11$\,\mathrm{\%}$  and the throughput was increased by 189.84$\,\mathrm{\%}$ compared to CPU execution,
and compared to the GPU, the power consumption was reduced by 89.72$\,\mathrm{\%}$ and the throughput was increased by 34.47$\,\mathrm{\%}$. 
These results demonstrate that the FPGA implementation with DPU is more power-efficient than with the CPU or GPU and can be applied to on-board processing systems with limited resources and power consumption.

While the above work used DPU for a single inference, our proposed method utilizes it for multiple inferences in a time-shared manner. 
We also implement multi-threading to improve the overall throughput by increasing the efficiency of DPU utilization.



\section{Proposed Method}
\label{fer}

\subsection{System overview}
The processing of the proposed system is divided into two steps, as shown in Fig. 1: (1) face detection and (2) facial expression recognition. 
The face detection step utilizes DenseBox, a DNN model, to detect faces by analyzing camera frames as input.
In the next step, a CNN facial expression recognition model is utilized to identify facial expressions within the region of interest. 
Both steps involve inference processes and are offloaded to the same DPU running on the FPGA. 
Pre-processing tasks (such as image resizing, cropping, and gray scaling) are performed by the CPU.

Here, we describe the DNN for face detection and facial expression recognition. 
We use the DenseBox model\nobreak\cite{denseface} provided by Xilinx as the face detection model. 
The inference with DenseBox is very lightweight, allowing fast detection even when run on a DPU with a low computing performance.
The Wider Face dataset\nobreak\cite{yang2016wider} was used to train this model.

\begin{figure}[t]
  \begin{center}
  \includegraphics[width=.95\columnwidth]{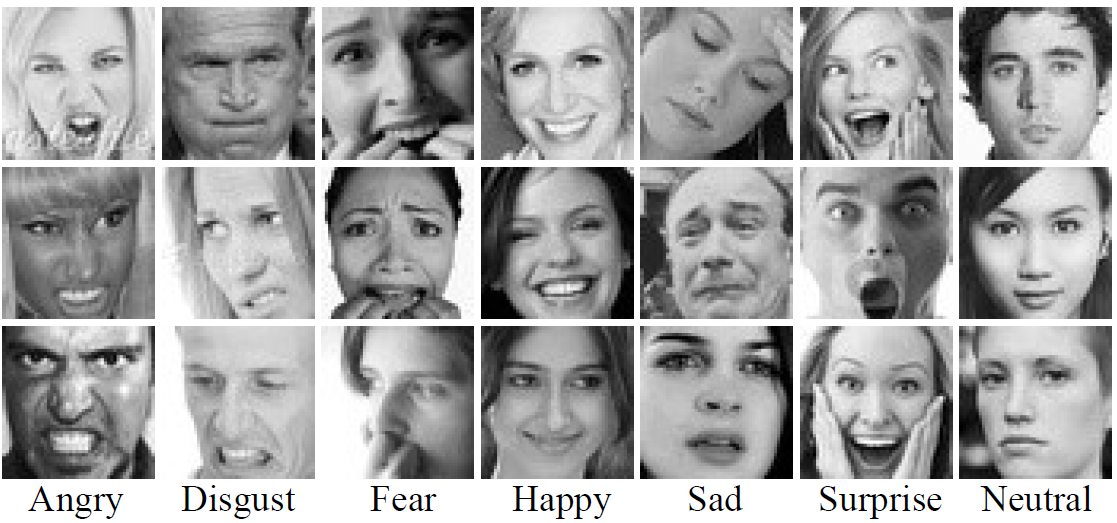}
  \caption{Sample images from FER-2013 dataset}
  \label{fig:fer_sample}
  \end{center}
\end{figure}

As for the facial expression recognition model, it was created using the network proposed by Guarniz et al\nobreak\cite{fercnn}. 
We used the FER-2013\nobreak\cite{goodfellow2013challenges} to train this model.
As shown in Fig.~\ref{fig:fer_sample}, this dataset consists of a total of 35,887 grayscale images of 48$\times$48~$\mathrm{px}$, with seven expression labels assigned~(Angry, Disgust, Fear, Happy, Sad, Surprise, and Neutral). 

\subsection{FPGA SoC architecture}
\begin{figure}[t]
\centerline{\includegraphics[width=.95\columnwidth]{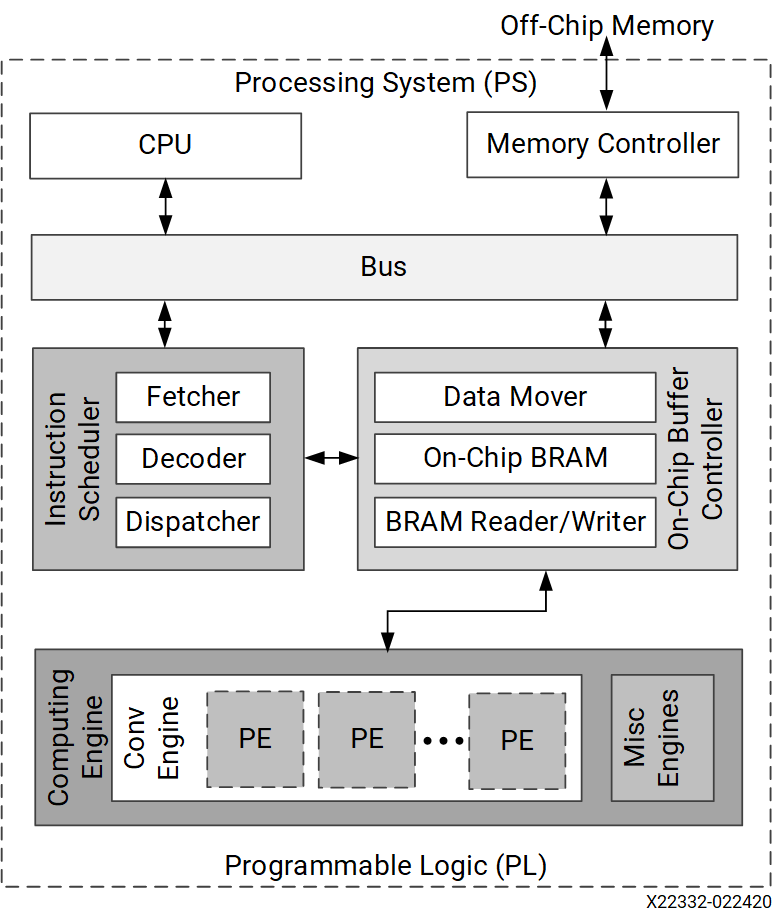}}
\caption{Hardware architecture\nobreak\cite{dpu}}
\label{fig:HW_arch}
\end{figure}

The system is implemented on a Xilinx Zynq Ultrascale+ MPSoC, which integrates the ARM Cortex-A53 processor and Xilinx’s UltraScale+ FPGA. 
We used a Xilinx Kria KV260 development board with this SoC as an evaluation board. Xilinx Vivado 2021.1 was used to create the hardware design incorporating the DPU. As shown in Fig.~\ref{fig:HW_arch}, 
the DPU is embedded on an FPGA part and connected to the CPU by an Advanced eXtensible Interface (AXI) bus. 
The CPU and DPU share off-chip memory, and the DPU fetches instructions from this memory to control the operation of the arithmetic engine. 
In contrast, the on-chip memory on the FPGA is used as a buffer for input images, neural network parameters (weights and biases), and intermediate features to achieve high throughput and efficiency. 
The processing element (PE) serves as the computing engine of the DPU and operates in a multi-stage pipeline to achieve a high computing performance.

\subsection{Offloading of inference to DPU}

\begin{figure*}[t]
    \centerline{\includegraphics[width=.95\textwidth]{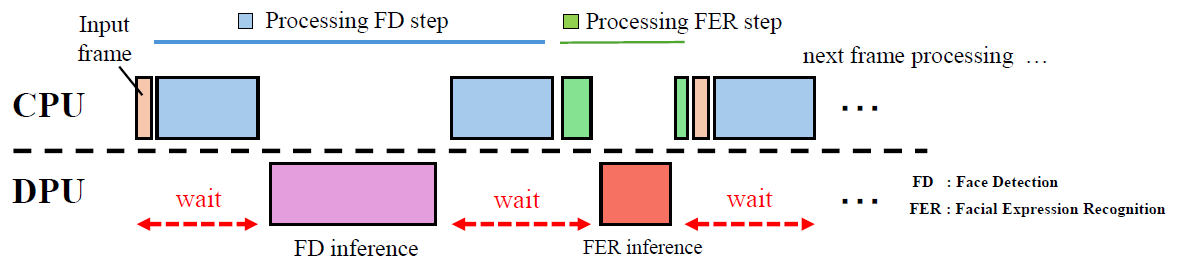}}
    \caption{Process flow of face detection and facial expression recognition by single thread}
    \label{fig:single}
  \end{figure*}

  \begin{figure*}[t]
    \centerline{\includegraphics[width=.95\textwidth]{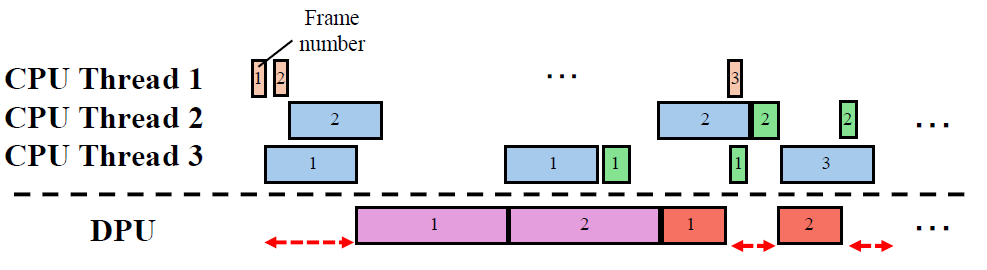}}
    \caption{Process flow of face detection and facial expression recognition by multi thread}
    \label{fig:multi}
  \end{figure*}
The DPU supports various architectures with different levels of arithmetic performance, 
allowing for their selection based on specific applications. 
The size of the DPU is determined by the parallelism of the convolutional units, offering a range of sizes from B512 to B4096 (512 and 4096 denote the number of operations executed per cycle). 
When a DPU with a high arithmetic performance is integrated into a circuit, the resource consumption of the FPGA increases. 
The face detection and facial expression recognition processes executed by this system are relatively lightweight compared to other DNN inference tasks, such as 3D object detection, so
the system is implemented on an FPGA part with a B512, which has the smallest circuit size.

The accelerator implemented in the previous work is a dedicated circuit type that directly circuits the DNN model for facial expression recognition through high-level synthesis.
Therefore, only the DNN inference for facial expression recognition can be performed, while a separate accelerator needs to be implemented for face detection.
In contrast, the DPU used in the proposed system is a systolic array type, which allows different types of inference to be run on the same accelerator.
In this work, we propose a method to efficiently utilize FPGA resources without increasing the number of accelerators by running face detection and facial expression recognition on the DPU in a time-shared manner.

To offload the DNN to the DPU, it is necessary to transform the model using the Vitis AI 2.5 tool.
Vitis AI is a DNN framework provided by Xilinx, which is a development environment capable of converting the constructed models into a format that can be run on the DPU.
The AI quantizer and AI compiler included in the Vitis AI tools were utilized to convert the DNN models.
The AI quantizer is a tool that quantizes $32\,\mathrm{bit}$ floating-point models into $8\,\mathrm{bit}$ fixed-integer models, maintaining as much accuracy as possible.
Thus, quantizing the model optimizes memory usage and reduces the number of hardware operations.
As for the AI compiler, it converts the models quantized to 8 bits by the AI quantizer into models that can be run on the DPU. These tools enable DNN models to be run on the DPU by quantizing and compiling the models.
\subsection{Multi-threading strategy}
The processing flow when this system is executed with a single thread is shown in Fig.~\ref{fig:single}.
While the CPU is processing, the DPU is forced to wait for the next process, which prevents their efficient usage. 
Therefore, multi-threading is implemented to improve the efficiency.
Our system uses Threading, a multi-threaded Python library, to execute face detection and facial expression recognition in different threads.
Fig.~\ref{fig:multi} shows the process flow in the case of multi-threaded execution.
As a precondition for multi-threading, we need to prepare one thread dedicated to camera input, since the camera input by OpenCV is not thread-safe. 
Therefore, a thread dedicated to camera input and two processing threads is created and executed with a total of three threads.
The two processing threads run on different frames and share one DPU.
Compared to Fig.~\ref{fig:single}, the multi-threaded implementation in Fig.~\ref{fig:multi} reduces the DPU latency by increasing the frequency at which the CPU instructs the DPU to execute.
This implementation method is expected to improve the efficiency of DPU utilization and the overall throughput compared to a single thread run.

\section{evaluation}
\label{ex_re}

\subsection{Inference for face detection}
We evaluate the face detection model run on the DPU in terms of average precision and latency and compare it with the performance of the previous work. 
For this evaluation, we use the AFW dataset\nobreak\cite{Zhu2012FaceDP}, which consists of 205 images of various resolutions and 473 annotated faces. 
These images include faces angled from the side and images with scale, illumination, occlusion, and other challenges.

The results of the evaluation of face detection using the AFW dataset are shown in Table~\ref{tab:fd}.
A comparison is made between the model run on the DPU and the Haar Cascade detector.
The input size of DenseBox is 640$\times$480~$\mathrm{px}$, which requires resizing the input image,but since the Haar Cascade detector performs inference without resizing, 
we ensured a fair comparison by adding the resize time to the DenseBox latency. 

The accuracy of the quantization model run on the DPU was 0.917, which is approximately 1.73 times higher than the accuracy of the method using the Haar Cascade detector in the previous work.
The latency of the model run on the DPU was $42.10\,\mathrm{ms}$, which is about 18.95 times shorter than the previous work.
These results confirm that the DNN model in the DPU outperforms the previous work in terms of both recognition accuracy and latency. 
 
\begin{table}[t]
    \centering
    \caption{Accuracy and latency results for face detection}
    \label{tab:fd}
    \scalebox{1.00}{
    \begin{tabular}{cc|cc}
        \hline \hline 
        \multicolumn{2}{c|}{Method}&  Average precision & Latency[ms] \\\hline 
       
        \multicolumn{1}{c|}{\multirow{2}{*}{CPU}}&Haar Cascade &  \multirow{2}{*}{0.531}& \multirow{2}{*}{798}\\ 
        \multicolumn{1}{c|}{}&(Previous work)&& \\  

        \multicolumn{1}{c|}{\multirow{2}{*}{DPU}}&\bf{DenseBox}  & \multirow{2}{*}{\bf{0.917}} & \multirow{2}{*}{\bf{42.10}} \\ 
        \multicolumn{1}{c|}{}&\bf{(Our work)}&& \\ \hline 

    \end{tabular}
    }
\end{table}

\subsection{Inference for facial expression recognition}
Next, we evaluate the facial expression recognition model run on the DPU in terms of recognition accuracy (percentage) and latency. 
For this evaluation, we use the FER-2013 dataset\nobreak\cite{goodfellow2013challenges}. 
The accuracy and latency of facial expression recognition for each DNN are shown in Table~\ref{tab:fer}.

The facial expression recognition model run on the DPU achieved an accuracy of $67.4\,\mathrm{\%}$ and a latency of $7.34\,\mathrm{ms}$.
The model in the previous work, which was offloaded to an FPGA, achieved an accuracy of $66\,\mathrm{\%}$ and latency of $6.36\,\mathrm{ms}$.
Despite our system surpassing the previous work in terms of facial expression recognition accuracy, its latency was lower.

\subsection{Overall system comparison}
Next, to investigate the effectiveness of the hardware configuration of our system using B512, we evaluate the FPGA resource consumption, system power consumption, and throughput for DPUs of different sizes.
Additionally, we evaluate the change in DPU utilization efficiency and processing performance brought about by multi-threading.Table~\ref{tab:throughput/power} shows the results of these comparisons.
However, since it is not possible to completely reproduce the implementation environment of the previous work, only its hardware configuration is shown here.
Specifically, we created a system that runs on the DPU for face detection using the Haar Cascade detector on the CPU and facial expression recognition using CNN.
For power consumption, we measured the difference between the maximum power consumption of the entire board when the system was idle ($7.8\,\mathrm{W}$) and when the system was running (hereinafter, “peak power”). 

First, we evaluate the FPGA resource consumption and throughput.
The previous work consumed 22,465 ALMs and 112 DSPs, while in contrast, our system consumed 27,023 LUTs and 118 DSPs.
When comparing the circuit sizes, we found that the previous work consumed fewer FPGA resources.
Although the circuit size of our system is larger than that of the previous work, it is still possible to run facial expression recognition and face detection inference on the same DPU.
The throughput of the previous work was $11.67\,\mathrm{FPS}$, whereas our achieved $14.69\,\mathrm{FPS}$  with one thread and $25.00\,\mathrm{FPS}$ with two threads.
As a real-time face detection and facial expression recognition system, this system achieves a sufficient throughput.

\begin{figure}[t]
    \begin{center}
    \includegraphics[width=.95\columnwidth]{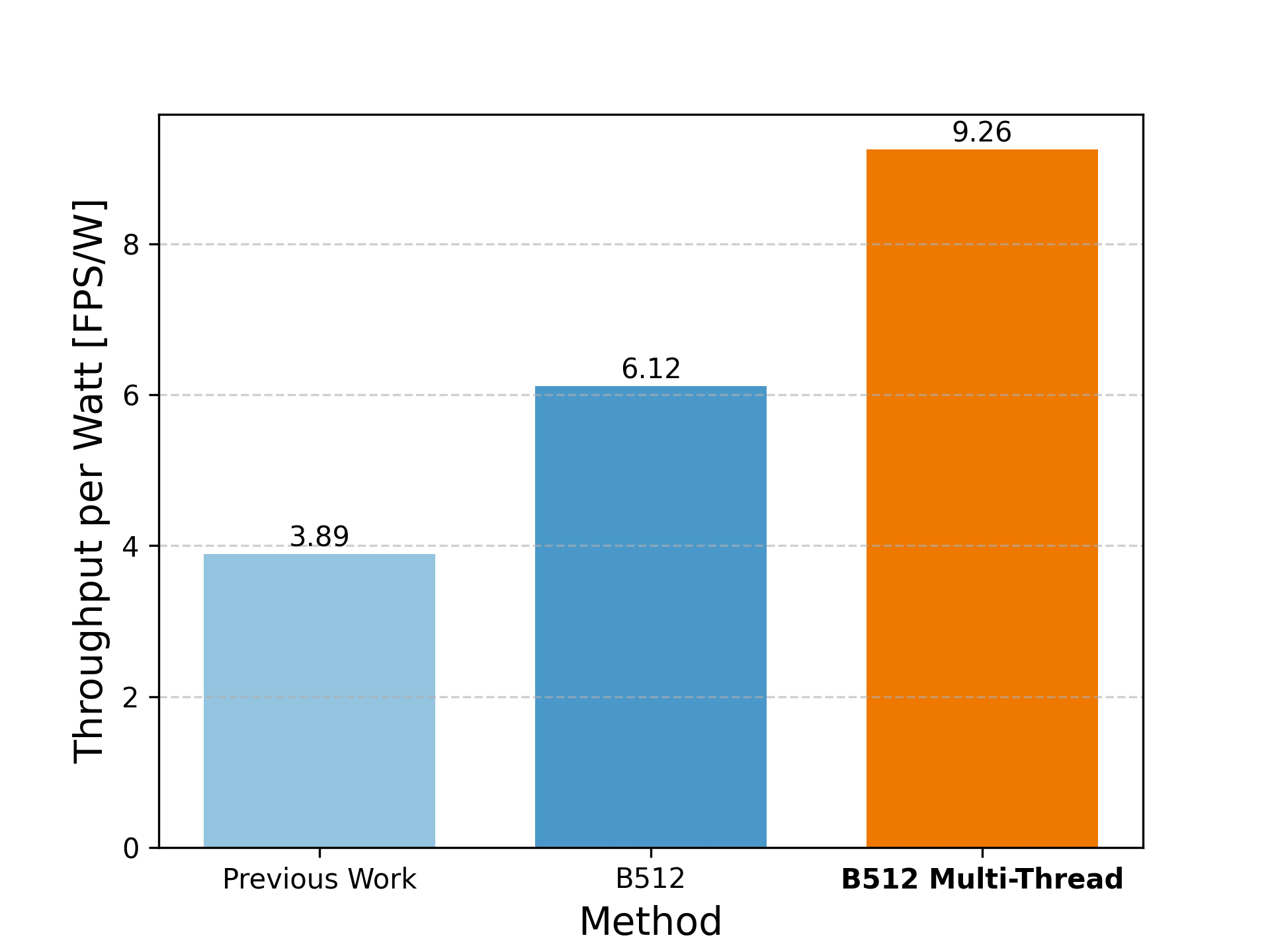}
    \caption{Throughput per power consumption for each system}
    \label{fig:bar_graph}
    \end{center}
\end{figure}

Next, we evaluate the power consumption and throughput.
The power consumption of the previous work was $2.3\,\mathrm{W}$, while that of our system was $2.4\,\mathrm{W}$ for one thread and $2.7\,\mathrm{W}$ for two threads.
Compared to the previous work, our system consumes slightly more power, but the throughput is about 2.14 times higher with two threads.
Fig.~\ref{fig:bar_graph} shows the throughput per power consumption, which for our system with B512 was $6.12\,\mathrm{FPS/W}$ for one thread and $9.26\,\mathrm{FPS/W}$ for two threads.
Compared to the previous work, the multi-threaded system improved by approximately 2.4 times, thus confirming the effectiveness of the system in terms of both power consumption and throughput.

Finally, we evaluate the efficiency of DPU utilization.
The DPU utilization rate in Table~\ref{tab:throughput/power} shows the ratio of the DPU execution time to the total system execution time.
Since the CPU and DPU can run in parallel, the higher the utilization ratio, the more efficiently the DPU is used.
With one thread, the total time for face detection and facial expression recognition was $22.85\,\mathrm{\%}$, and with two threads, it was $78.44\,\mathrm{\%}$.
These results demonstrate that the DPU utilization efficiency by multi-threading is high and contributes to the overall system throughput improvement. 

\begin{table}[t]
    \centering
    \caption{Accuracy and latency results for expression recognition}
    \label{tab:fer}
    \scalebox{1.00}{
    \begin{tabular}{c|ccc}
        \hline \hline 
        Method& Accuracy[\%] & Latency[ms]   \\ \hline 
        \multirow{2}{*}{CNN(Previous work)}&\multirow{2}{*}{66}&\multirow{2}{*}{6.36}\\ 
        && \\
        \bf{CNN(Our wor)}    & \bf{67.4}& \bf{7.34} \\  \hline

    \end{tabular} 
    }
\end{table}
\begingroup
\renewcommand{\arraystretch}{1.1}
\setlength{\tabcolsep}{5pt} 
\begin{table*}[h]
    \centering
    
    \caption{Comparison of FPGA resources, throughput and power consumption with previous work and DPU utilization}
    \label{tab:throughput/power}
    
    \scalebox{1.0}{
    \begin{tabular}{c|ccc|cccccccc}
    
        \hline \hline 
        \multirow{2}{*}{Method}& \multicolumn{3}{c|}{FPGA resource} &\multirow{2}{*}{Thread}   & \multirow{2}{*}{Throughput [FPS] }&\multirow{2}{*}{Peak power [W]} &\multirow{2}{*}{Power [W]} & \multicolumn{3}{c}{DPU utilization [\%]} \\ \cline{2-4}
  & ALMs or LUTs *& DSPs& BRAMs&&&&&FD&FER&Total\\ \hline
  Haar Cascade&\multirow{2}{*}{34,593}&\multirow{2}{*}{230}&\multirow{2}{*}{44}&\multirow{2}{*}{1}& \multirow{2}{*}{11.67}&\multirow{2}{*}{10.1}&\multirow{2}{*}{2.3}&\multirow{2}{*}{-}&\multirow{2}{*}{-}&\multirow{2}{*}{-}\\
  (Previous work)& & & && &&&\\
  \multirow{2}{*}{\bf{Our work}}  &\multirow{2}{*}{\bf{27,023}} &\multirow{2}{*}{\bf{118}}&\multirow{2}{*}{\bf{12}}&1 &14.69&10.2&2.4&15.84&7.01&22.85 \\ 
    & & & &\bf{2} &\bf{25.00}&\bf{10.5}&\bf{2.7}&\bf{49.72}&\bf{28.72}&\bf{78.44}\\ \hline
    \end{tabular}
    }
    \begin{tablenotes}
        \scriptsize
        \item[1] (*The previous work utilizes ALMs because the boards used are Intel boards.)
        \item[2](FD: Face Detection, FER: Facial Expression Recognition) 
    \end{tablenotes}
\end{table*}
\endgroup

\section{discussion}

\label{discussion}

\begin{table}[t]
    \centering
    \caption{Throughput and power consumption by frequency}
    \begin{tabular}{cccccc}
        \hline \hline 
        \multirow{2}{*}{Frequency} & Throughput & Peak Power & Power & Throughput/Power \\
        & [FPS] & [W] & [W] & [FPS / W] \\
        \midrule
        600 & 27.78 &  11.7 & 2.3 & 12.08 \\
        500 & 25.00 &  11.6 & 2.2 & 11.36 \\
        \bf{400 }& \bf{25.00} & \bf{11.4} &\bf{2.0} &\bf{12.50} \\
        300 & 21.74 &  11.2 & 1.8 & 12.08 \\
        \bottomrule
    \end{tabular}
    \label{tab:frequency}
\end{table}


\begingroup
\renewcommand{\arraystretch}{1.1} 
\begin{table}[tb]
    \centering

    \caption{Comparison of FPGA resources and performance with different DPU sizes}
    \begin{tabular}{c|ccc|cccccc}
        \hline \hline 
        \multirow{2}{*}{Size} &\multicolumn{3}{c|}{FPGA resource}&\multirow{2}{*}{Thread} &Throughput  & Power \\\cline{2-4}
             & LUTs & DSPs & BRAMs & &   [FPS]     &   [W]    \\
             \hline
        \multirow{2}{*}{\bf{512}}& \multirow{2}{*}{\bf{27,023}} & \multirow{2}{*}{\bf{118}} & \multirow{2}{*}{\bf{12.0}}  &1& 14.69 &2.4  \\
            &        &     &  &\bf{2}& \bf{25.00} & \bf{2.7} \\
        \multirow{2}{*}{1024}& \multirow{2}{*}{34,593} & \multirow{2}{*}{230} & \multirow{2}{*}{44.0}  &1& 19.21  & 2.5  \\
             &        &     &   &2& 27.74  & 3.1  \\
        \multirow{2}{*}{2034}& \multirow{2}{*}{41,861} & \multirow{2}{*}{438} & \multirow{2}{*}{60.5}& 1& 20.80 &2.9 \\
              &        &     &    & 2& 27.75  & 3.2  \\  
        \multirow{2}{*}{4096}& \multirow{2}{*}{51,561} & \multirow{2}{*}{710} & \multirow{2}{*}{82.5} &1& 23.77 & 3.2 \\
              &        &     &    & 2& 27.74 & 3.5 \\
              \hline
    \end{tabular}
    \label{tab:other_dpu}
\end{table}
\endgroup
\begin{figure}[t]
    \begin{center}
    \includegraphics[width=.95\columnwidth]{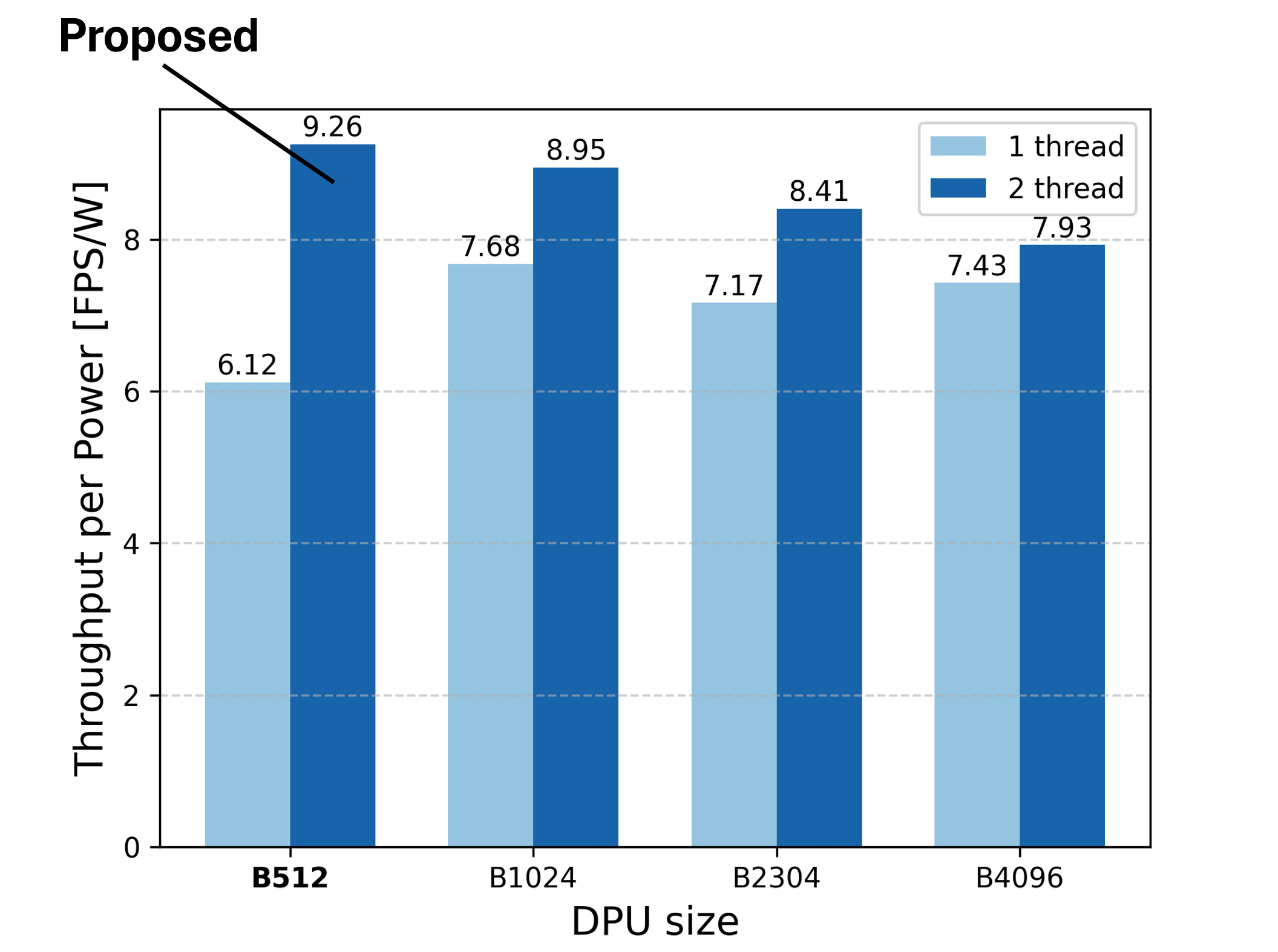}
    \caption{Throughput per power consumption for hardware with each DPU embedded}
    \label{fig:bar_graph2}
    \end{center}
\end{figure}
\subsection{Analysis of optimum operating frequency}
DNN inference on the system is run on the DPU at an operating frequency of $400\,\mathrm{MHz}$.
Considering that the operating frequency has a notable influence on processing performance and power consumption, we investigate the optimal DPU operating frequency.
In this subsection, we clarify the optimal run on the DPU in terms of throughput, power consumption, and DPU utilization efficiency.

Table~\ref{tab:frequency} shows the throughput and power consumption of the B512 multi-threaded runs at different frequencies.
The best throughput was $27.78\,\mathrm{FPS}$ at $600\,\mathrm{MHz}$, but this also had the highest power consumption. 
Thus, there was a trade-off between throughput and power consumption.
However, even if the operating frequency was increased above $400\,\mathrm{MHz}$, the rate of improvement in throughput was low.
This phenomenon may stem from a processing bottleneck within the CPU.
Consequently, even if the inference in the DPU on the FPGA was accelerated, the overall throughput would not experience significant enhancement. 
The optimum throughput per power consumption ratio was revealed at $400\,\mathrm{MHz}$.
Furthermore, given that a throughput of $25.00\,\mathrm{FPS}$ suffices, we conclude that the optimal operating frequency is $400\,\mathrm{MHz}$.


\subsection{Comparison with other sizes of DPU}
Above, we evaluated our system on an FPGA incorporating B512. 
In this section, we investigate which DPU size is most effective in terms of FPGA resource usage, throughput, and power consumption for the entire system.
A comparison is made between the B512-integrated FPGA and other DPU sizes in our system.
Table~\ref{tab:other_dpu} shows the FPGA resource usage, throughput, and power consumption for DPU sizes B1024, B2034, and B4096 with an operating frequency of $400\,\mathrm{MHz}$.

When comparing the FPGA resources used with the incorporation of DPU, we observed that the B512 used in our system exhibited the lowest computational performance, resulting in a smaller circuit size compared to the others. 
Moreover, as the size of the DPU increased, there was a notable increase in the usage of DSP and BRAM.
Comparing B4096 to B512, the LUT increased by approximately 1.9 times, which signifies a small increase. 
However, the DSP has increased by approximately 6.0 times and the BRAM by approximately 6.9 times.
The fact that the BRAM increased this much indicates that it consumes a significantly large amount of the FPGA resources.

When run on a single thread, increasing the size of the DPU increased the throughput, but only by approximately 1.7 times when the size was changed from B512 to B4096.
The increase in throughput became even more marginal when the system was operated with multiple threads. Specifically, there was hardly any alteration between B1024 and B4096.
We attribute this phenomenon to bottlenecks in the CPU processes, apart from the FPGA-based inference process. 
In short, the power consumption rises as the DPU size increases.

Fig.\ref{fig:bar_graph2} shows the throughput per power consumption for each DPU size and thread.
Here, the proposed B512 multi-threaded execution stands out as the most efficient at $9.26\,\mathrm{FPS/W}$.
On the basis of this finding, we conclude that the B512 multi-threaded execution is the most effective due to its optimal balance between processing performance and FPGA resource usage.

\section{Conclusion}
\label{conclusion}
In this work, we implemented a facial expression recognition system using DPU, a general-purpose CNN accelerator, 
and applied a systolic array accelerator to run the inference of two DNNs for face detection and facial expression recognition on the same DPU in a time-shared manner. 
We also proposed a multi-threaded system to improve the overall throughput while increasing the efficiency of DPU utilization.
Compared to a previous method using the Haar Cascade detector, the face detection using our approach achieved an accuracy improvement of approximately 1.73 times and a latency reduction of approximately 18.95 times. 
Moreover, the accuracy of facial expression recognition was $67.4\,\mathrm{\%}$ and the latency per image was just $7.34\,\mathrm{ms}$.

Although the circuit size in our system is slightly larger than in the previous work, it is possible to run facial expression recognition and face detection inference on the same DPU. 
Furthermore, in multi-threading execution, the throughput per power consumption was improved by 2.4 times compared to the previous work, achieving a throughput of 25 FPS.
Therefore, we conclude that the hardware configuration utilizing multi-threading with the same DPU achieves superior results compared to the previous work while maintaining a small circuit size.

Future works include reducing the power consumption even further while maintaining sufficient throughput. 
We also plan to optimize the processing for actual operation, such as performing face detection once every few frames, given that the position of the face remains unchanged.




\bibliographystyle{myjunsrt.bst}
\bibliography{reference}

\end{CJK}
\end{document}